\documentclass[10pt,journal,compsoc]{IEEEtran}
\usepackage[noadjust]{cite}

\usepackage{multirow,multicol}

\usepackage[tbtags]{amsmath}
\usepackage{amsbsy}
\usepackage{amssymb}
\usepackage{amsfonts}
\usepackage{bbm}

\usepackage{graphicx}
\usepackage[caption=false]{subfig}
\usepackage{url}
\usepackage{textcomp}
\usepackage{algorithmic}
\usepackage{algorithm}
\usepackage{balance}

\usepackage{array}
\usepackage{hyperref}
\usepackage{color}

{\begin{list}               %    with flush left bullets.
    {$\bullet$ \hfill}{
        \setlength{\leftmargin}{\parindent}
        \setlength{\parsep}{0.04\baselineskip}
        \setlength{\itemsep}{0.5\parsep}
        \setlength{\labelwidth}{\leftmargin}
        \setlength{\labelsep}{0em}}
    }
{\end{list}}

\providecommand{\eref}[1]{\eqref{#1}}  % call \eqref from amstex
\providecommand{\cref}[1]{Chapter~\ref{#1}}
\providecommand{\sref}[1]{Section~\ref{#1}}
\providecommand{\fref}[1]{Fig.~\ref{#1}}

\renewcommand{\vec}[1]{\ensuremath{\boldsymbol{#1}}}

\providecommand{\w}{\omega}

\graphicspath{{figs/}}
\DeclareGraphicsExtensions{.jpg, .png}

\begin{document}

\title{Lossless White Balance For Improved Lossless CFA Image and Video Compression}

\author{Yeejin~Lee,~\IEEEmembership{Member,~IEEE,}
        and~Keigo~Hirakawa,~\IEEEmembership{Senior~Member,~IEEE}% <-this 
\IEEEcompsocitemizethanks{\IEEEcompsocthanksitem Y.~Lee is with the Department of Electrical and Information Engineering, Seoul National University of Science and Technology, Seoul, South Korea, mail: yeejinlee@seoultech.ac.kr.\protect\\
E-mail: see http://www.michaelshell.org/contact.html
\IEEEcompsocthanksitem K.~Hirakawa is with the Department of Electrical and Computer Engineering, University of Dayton, OH, 45469 USA E-mail: khirakawa1@udayton.edu }}% <-this % stops an unwanted space

\IEEEtitleabstractindextext{%
\begin{abstract}
Color filter array is spatial multiplexing of pixel-sized filters placed over pixel detectors in camera sensors. The state-of-the-art lossless coding techniques of raw sensor data captured by such sensors leverage spatial or cross-color correlation using lifting schemes. In this paper, we propose a lifting-based lossless white balance algorithm. When applied to the raw sensor data, the spatial bandwidth of the implied chrominance signals decreases. We propose to use this white balance as a pre-processing step to lossless CFA subsampled image/video compression, improving the overall coding efficiency of the raw sensor data.
\end{abstract}

\begin{IEEEkeywords}
Color filter array, image compression, video compression, lifting scheme, white balance.
\end{IEEEkeywords}}

\maketitle

\IEEEdisplaynontitleabstractindextext

\IEEEraisesectionheading{\section{Introduction}}

\IEEEPARstart{P}{hotographers} and videographers that require the digital representation of the highest image or video quality need access to raw sensor data. Raw sensor data affords maximum technical and creative flexibility to post-capture processing, steps of which typically include nonlinearity correction, demosaicking, color correction, white balance, denoising, color grading, tone mapping, and gamma correction. A modern camera records at 10 to 14 bit data per pixel, making it a very data-intensive workflow with challenging storage capacity constraints. This is particularly true of video signals, especially given that the modern camera hardware yields high resolution (8K~(33 megapixels) or higher) and high framerate (60 Hz or faster).

\begin{figure}[t]
    \centering
    \begin{tabular}{@{}c@{~}c@{~}c@{}}
        \includegraphics[width=0.16\textwidth]{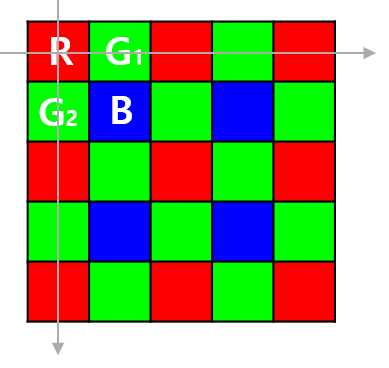}&
        \includegraphics[width=0.16\textwidth]{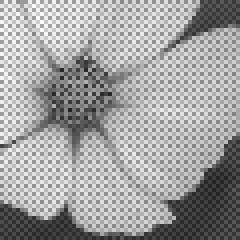}&
        \includegraphics[width=0.16\textwidth]{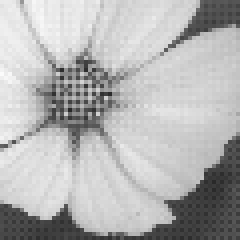}\\
        (a)&(b)&(c)
    \end{tabular}
    \caption{(a) Bayer color filter array pattern~\cite{Bayer75}. (b) Zoomed portion of a raw sensor image. (c) Same zoomed raw sensor image with the proposed lossless white balance applied.}
    \label{fig:bayerPattern}
\end{figure}

Compression of raw sensor data differs from the conventional image and video compression, in the sense that the most single-chip image and video sensors record color values using an array of pixel-sized color filters called color filter array (CFA). \fref{fig:bayerPattern}(a) shows a Bayer CFA pattern that spatially multiplexes red-, green-, and blue-colored absorptive filters placed over an array of pixel sensors. Because each pixel sensor captures only one of the three color components, the raw data captured by such sensor configuration is a ``CFA subsampled'' version of the full-color image or video frame. Neighboring pixels record alternating color components, making it difficult to exploit spatial correlations in compression. See \fref{fig:bayerPattern}(b).

High resolution, high framerate raw video sensor data are considerably noisier than raw images from digital single-lens reflex (DSLR) still images. Unlike non-raw compression which operates on data already processed by camera's post-capture processing~(most likely including a color balance and an image denoising steps), raw compression means no color balancing and denoising have been applied. This implies that the differences in the color intensity levels and noise levels of still and video raw data are even more pronounced in raw data compression. Consequently, raw video data is much more challenging to compress than raw still images. The low bitdepths, lower resolution, higher noise, and lack of color balance are less favorable for compression, especially losslessly. 

\begin{figure*}[t]
    \centering
    \includegraphics[scale=0.55]{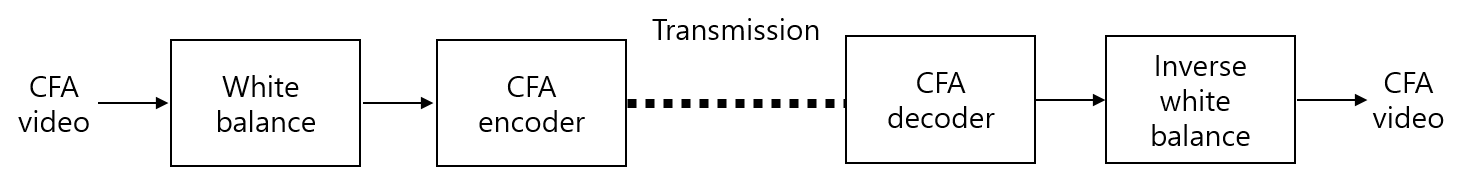}
    \caption{The proposed compression pipeline. The proposed lossless white balance is applied as a pre-processing step to an existing CFA subsampled image/video compression. In decoding, the white balance is inverted following the CFA decoder.}
    \label{fig:system}
\end{figure*}

In this paper, we propose a novel lossless white balance technique using the scalar multiple lifting scheme. This white balance procedure is designed to restore the balance between red, green, and blue channels in raw sensor data affected by illumination color, detector's spectral quantum efficiency, and color filter transmittance (see \fref{fig:bayerPattern}(c)). The proposed white balance on its own does not penalize nor benefit the coding efficiency. But when this white balance is used as a pre-processing step to lossless CFA compression algorithms (see \fref{fig:system}), the overall coding efficiency of state-of-the-art CFA compression improves by a decisive margin. The proposed white balance technique is implemented as an integer operation, making it an exactly invertible process that is ideal for lossless compression with no risk of irreversible rounding effects.

The remainder of this paper is organized as follows. In \sref{sec:background}, we provide a review of CFA image/video compression and background on lifting. In \sref{sec:cfaSampling}, we offer the theoretical analysis of CFA subsampling and empirical evidence of white balance reducing the spatial bandwidth of raw sensor data. The proposed lossless white balance method is introduced in \sref{sec:proposed}. We demonstrate the compression gain through experimental verification in \sref{sec:results} before making concluding remarks in \sref{sec:conclusion}. In the remainder of this paper, we do not draw the distinction between CFA image and CFA video frame because the proposed lossless white balance is agnostic to them. As a convention, the exposition below assumes that the red filter is at the origin (see \fref{fig:bayerPattern}(a)), but the analyses and techniques developed in this paper are equally valid to shifted versions of Bayer CFA pattern.

% In \sref{sec:method}, we develop the proposed transform using lifting scheme for CFA sampled videos. Experimental results shown in \sref{sec:results} are conducted using camera hardware configurations matching a typical video acquisition setup. Concluding remark are made in \sref{sec:conclusion}.

\section{Background and Review}
\label{sec:background}
\subsection{CFA Subsampled Image/Video Compression}
\label{sec:cfa_compression}

The CFA subsampled sensor data undergoes a series of image and video processing steps to render a full-color image. These steps---commonly referred to as ISP (image signal processing)---include nonlinearity correction, demosaicking, color correction, white balance, and gamma correction. In particular, demosaicking is an interpolation process that reconstructs the unobserved tristimulus color components based on surrounding pixels. Although demosaicking is a necessary operation for digital cameras, it triples the data rate relative to the raw sensor values, which is inefficient for image compression.

Image compression before demosaicking has been sought after as a potential alternative to the conventional post-demosaicking compression because of the reduced number of pixel color components that need to be encoded. A handful of methods have been developed for this purpose. The error introduced by lossy compression of raw sensor data can be amplified by the subsequent ISP steps such as color correction, white balance, and gamma correction~\cite{Lee17}. Indeed, methods such as~\cite{Lin16, Doutre08} that rearrange data similar to $4:2:2$ chroma subsampling incurs penalties that are numerically small in terms of raw sensor data reconstruction, but nevertheless, yield poor post-ISP results~\cite{Kim15}. Previously, we developed a Camera Aware Multi-Resolution Analysis (CAMRA) for CFA subsampled data compression, aimed at minimizing the amplification by ISP of error introduced to the lossy reconstruction of raw sensor data.

By contrast, lossless raw sensor data compression requires a considerably higher number of bits to encode. Yet, applications such as broadcasting, cinema, and medical imaging require the highest fidelity in a digital representation of the observed image and video signals, and so eliminating error from reconstruction (and subsequent amplification by ISP) is highly desirable. Various techniques aimed at exploiting or decorrelating cross-color correlation in a lossless manner have shown some promise~\cite{Malvar12, Suzuki2020, Kim14, Richter2019Bayer, Richter2019Rate}. The method in \cite{Zhang06} applied Mallat wavelet transform on CFA subsampled images~\cite{Mallat98}, which was shown by later work to implicitly perform luminance-chrominance decomposition in a manner similar to the standard color video coding schemes~\cite{hirakwa07framework, Korneliussen14, lee2018camera}. In our previous work, we exploited this decomposition to decorrelate two wavelet subbands that share the same chrominance components to yield state-of-the-art lossless CFA subsampled compression performance~\cite{lee2018camera,lee2020shift}.

\subsection{Lifting}
\label{sec:lifting}

\begin{figure*}[t]
\begin{minipage}{0.49\linewidth}
    \centering
    \centerline{\includegraphics[scale=0.55]{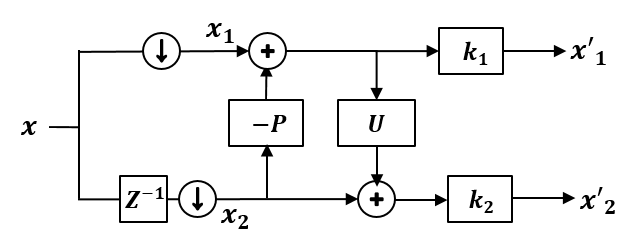}}
    \centerline{\footnotesize{(a)}}
\end{minipage}
\begin{minipage}{0.49\linewidth}
    \centering
    \centerline{\includegraphics[scale=0.55]{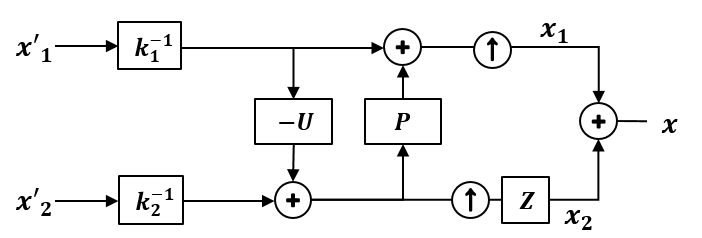}}
    \centerline{\footnotesize{(b)}}
\end{minipage}
\caption{Lifting based discrete wavelet transform. (a) Forward transform. (b) Inverse transform. Up and down arrows stand for upsampling and downsampling by 2, respectively, and delay lines are shown as $Z^{-1}$ and $Z$.}
\label{fig:liftingDWT}
\end{figure*}

Lifting is a reversible integer-to-integer decorrelation transform frequently used in wavelet transform and compression~\cite{Strang96, Sweldens98}. There are three canonical steps in lifting: split, predict, and update. As illustrated in \fref{fig:liftingDWT}(a), the split step (also known as lazy wavelet transform) deinterleaves pixel values into disjoint odd-indexed and even-indexed samples. The predict step~(denoted ``$P$'') then replaces the odd samples by the residual of predicting from the even samples (often referred to as ``detail coefficients''). Finally, the update step~(denoted ``$U$'') computes the approximation coefficients based on the predicted components and even samples. 
The polyphase representation of the lifting scheme can be factorized into a matrix form, as follows:
\begin{align}\label{eq:forwardLifting}
\left[ \begin{array}{c} x_1^\prime \\ x_2^\prime \end{array} \right] =  
\left[ \begin{array}{cc} k_1 & 0 \\ 0 & k_2 \end{array}\right]  
\left[ \begin{array}{cc} 1 & U \\ 0 & 1 \end{array}\right] 
\left[ \begin{array}{cc} 1 & 0 \\ -P & 1 \end{array}\right] 
\left[ \begin{array}{c} x_1 \\ x_2 \end{array} \right],
\end{align}
where $k_1$ and $k_2$ are the odd-sample scaling coefficient and the even-sample coefficient, respectively. For instance, lifting implementation of LeGall 5/3 wavelet transform commonly used in lossless compression is implemented by letting $P$ and and $U$ steps be integer convolution filtering with impulse response of $[0,\frac{-1}{2},\frac{-1}{2}]$ and $[\frac{1}{4},\frac{1}{4},0]$, respectively. The lifting scheme of \eref{eq:forwardLifting} is inverted by reversing the order of split/predict/update steps as described in \fref{fig:liftingDWT}(b). The polyphase representation of the inverse transform is represented as
\begin{align}\label{eq:reverseLifting}
\left[ \begin{array}{c} x_1 \\ x_2 \end{array} \right] =  
\left[ \begin{array}{cc} 1 & P \\ 0 & 1 \end{array}\right]
\left[ \begin{array}{cc} 1 & 0 \\ -U & 1 \end{array}\right] 
\left[ \begin{array}{cc} k_1^{-1} & 0 \\ 0 & k_2^{-1} \end{array}\right]  
\left[ \begin{array}{c} x_1^\prime \\ x_2^\prime \end{array} \right]. 
\end{align}

In practice, we would like the forward \eqref{eq:forwardLifting} and reverse \eqref{eq:reverseLifting} lifting schemes to be perfectly invertible---not just mathematically, but also free of loss of precision due to rounding errors in fixed point or floating operations. Lifting scheme can accomplish this by (i) setting $k_1=k_2=1$, and (ii) carrying out the predict and update steps as integer operations. For example, we round the prediction value $P\cdot x_e$ to perform the prediction step, as follows:
\begin{align}
    \left[ \begin{array}{cc} 1 & 0 \\ -P & 1 \end{array}\right] 
\left[ \begin{array}{c} x_1 \\ x_2 \end{array} \right]
=
\left[ \begin{array}{c} x_1 \\ x_2 - \lfloor P \cdot x_1\rfloor \end{array} \right].
\end{align}

\section{Motivation: CFA Sampling and White Balance} \label{sec:cfaSampling}
% Fig 2

In this section, we provide a detailed characterization of interactions between CFA sampling and white balance, based primarily on analysis in \cite{Hirakawa11, Korneliussen14} which drew inspirations from earlier works in \cite{glotzbach2001method, Alleysson05, Dubois05, hirakawa08spatio}. In particular, we will show that white balance has the desirable effect of decorrelating signal content within the CFA sampled image, resulting in increased coding efficiency.

\subsection{CFA Sampling}
Let $L(\lambda)$ be the spectrum of the illumination light, where $\lambda$ denotes spectral wavelength. Then the light observed at the image plane of the camera at pixel location $(i,j)$ is $L(\lambda)R(i,j\lambda)$, where $R(i,j,\lambda)$ is the spectral reflectance of the object in the scene. The recorded sensor value $\vec{x}(i,j)$---often referred to as the ``raw sensor data''---of a color frame at the pixel location $(i,j)$ is~\cite{Gijsenij11, Korneliussen14}
\begin{align} \label{eq:spectralResponse}
\begin{split}
x(i,j) =& \int C(i,j, \lambda) R(i,j,\lambda) d\lambda\\
=&\vec{c}(i,j)^T
\underbrace{\int 
\begin{bmatrix}\bar{r}(\lambda)\\\bar{g}(\lambda)\\\bar{b}(\lambda)\end{bmatrix}
R(i,j,\lambda)d\lambda}_{\vec{y}(i,j)=[y_r(i,j),y_g(i,j),y_b(i,j)]^T}.
\end{split}
\end{align}
Here, $C(i,j,\lambda)= \vec{c}(i,j)^T\left[ \bar{r}(\lambda),\bar{g}(\lambda),\bar{b}(\lambda)\right]^T$ is the spectral transmittance of CFA in the visible spectrum range (about 380-720nm), and $\vec{c}(i,j) = \left[ c_r(i,j),c_g(i,j),c_b(i,j)\right]^T\in \left\{ 0, 1\right\}^3$ is a sampling lattice at $(i,j)$~(e.g. $\vec{c}(i,j) = [1\:\:0\:\:0]^T$ is a red pixel). In other words, the recorded value $x(i,j)$ is a spatially subsampled version of the complete color image $\vec{y}(i,j)=[y_r(i,j),y_g(i,j),y_b(i,j)]^T$.

Consider the special case that the spatial arrangement of the CFA $\vec{c}(i,j)$ is a Bayer pattern. We may rewrite \eqref{eq:spectralResponse} as an inner product of the following form \cite{hirakawa08spatio,Alleysson05,Dubois05}:
\begin{align}\label{eq:cfaSampling}
&x(i,j) = \vec{c}(i,j)^T\vec{y}(i,j)\\
&=\vec{c}(i,j)^T
\begin{bmatrix} 1&2&1\\1&0&-1\\1&-2&1 \end{bmatrix}
\begin{bmatrix} 1&2&1\\1&0&-1\\1&-2&1 \end{bmatrix}^{-1}
%\begin{bmatrix} 1/4&1/2&1/4\\1/4&0&-1/4\\1/4&-1/2&1/4 \end{bmatrix}
\vec{y}(i,j)\notag\\
%&=\vec{c}(i,j)^T
%\begin{bmatrix} 1&2&1\\1&0&-1\\1&-2&1 \end{bmatrix}
%\begin{bmatrix} y_{\mu}(i,j) \\ y_{\gamma}(i,j)\\ y_{\beta}(i,j) \end{bmatrix}\notag\\
&= y_{\mu}(i,j) + ((-1)^i+(-1)^j)y_{\gamma}(i,j) + (-1)^{i+j}y_{\beta}(i,j)
\end{align}
where 
\begin{align}\label{eq:opponentcolor}
\begin{split}
y_{\mu}(i,j)=&\frac{y_r(i,j)}{4}+\frac{y_g(i,j)}{2}+\frac{y_b(i,j)}{4}\\
y_{\gamma}(i,j)=&\frac{y_r(i,j)}{4}-\frac{y_b(i,j)}{4}\\
y_{\beta}(i,j)=&\frac{y_r(i,j)}{4}-\frac{y_g(i,j)}{2}+\frac{y_b(i,j)}{4}.
\end{split}
\end{align}
In the demosaicking literature, $y_{\mu}(i,j)$ is thought to play a role of luminance, and $y_{\gamma}(i,j)$ and $y_{\beta}(i,j)$ are proxies for chrominance~\cite{Alleysson05}. Thus, the CFA sampled frame in \eqref{eq:cfaSampling} is a linear combination of completely observed luminance component and ``spatial frequency modulated'' chrominance components---where $(-1)^i$, $(-1)^j$, and $(-1)^{i+j}$ play the role of carrier frequency in amplitude modulation. Indeed, modulation at $(\pi,0)$,  $(0,\pi)$, and $(\pi,\pi)$ can be evidenced by the Fourier magnitude of $x$ in \fref{fig:fft}(g).

Demosaicking can therefore be reinterpreted as a demodulation task, recovering the luminance ($y_{\mu}(i,j)$) and chrominance ($y_{\gamma}(i,j)$ and $y_{\beta}(i,j)$) signals from CFA sampled data $x(i,j)$ (instead of estimating $[y_r(i,j),y_g(i,j),y_b(i,j)]^T$ directly)~\cite{hirakwa07framework,Korneliussen14, Alleysson05,Dubois05}. Once $y_{\mu}(i,j),y_{\gamma}(i,j),y_{\beta}(i,j)$ have been estimated at every pixel, then red, green, and blue signals are subsequently reconstructed by inverting the color transformation:
\begin{align}
    \begin{bmatrix}
        y_r(i,j)\\
        y_g(i,j)\\
        y_b(i,j)
    \end{bmatrix}
=
\begin{bmatrix}
1&2&1\\1&0&-1\\1&-2&1
\end{bmatrix}
\begin{bmatrix}
    y_{\mu}(i,j)\\
    y_{\gamma}(i,j)\\
    y_{\beta}(i,j)
\end{bmatrix}
.
\end{align}
The effectiveness of such demodulation task depends on the spatial bandwidth of the chrominance signals. As shown by \fref{fig:fft}(g), increased spatial bandwidth of $y_{\gamma}(i,j)$ and $y_{\beta}(i,j)$ would worsen the risk of aliasing, making it difficult to decouple from $y_{\mu}(i,j)$.

\subsection{White Balance}
\label{sec:whiteBalance}

\begin{figure*}
\centering
\begin{tabular}{@{}c@{~}c@{~}c@{~}c@{~}c@{}}
\includegraphics[width=.16\textwidth,height=.16\textwidth]{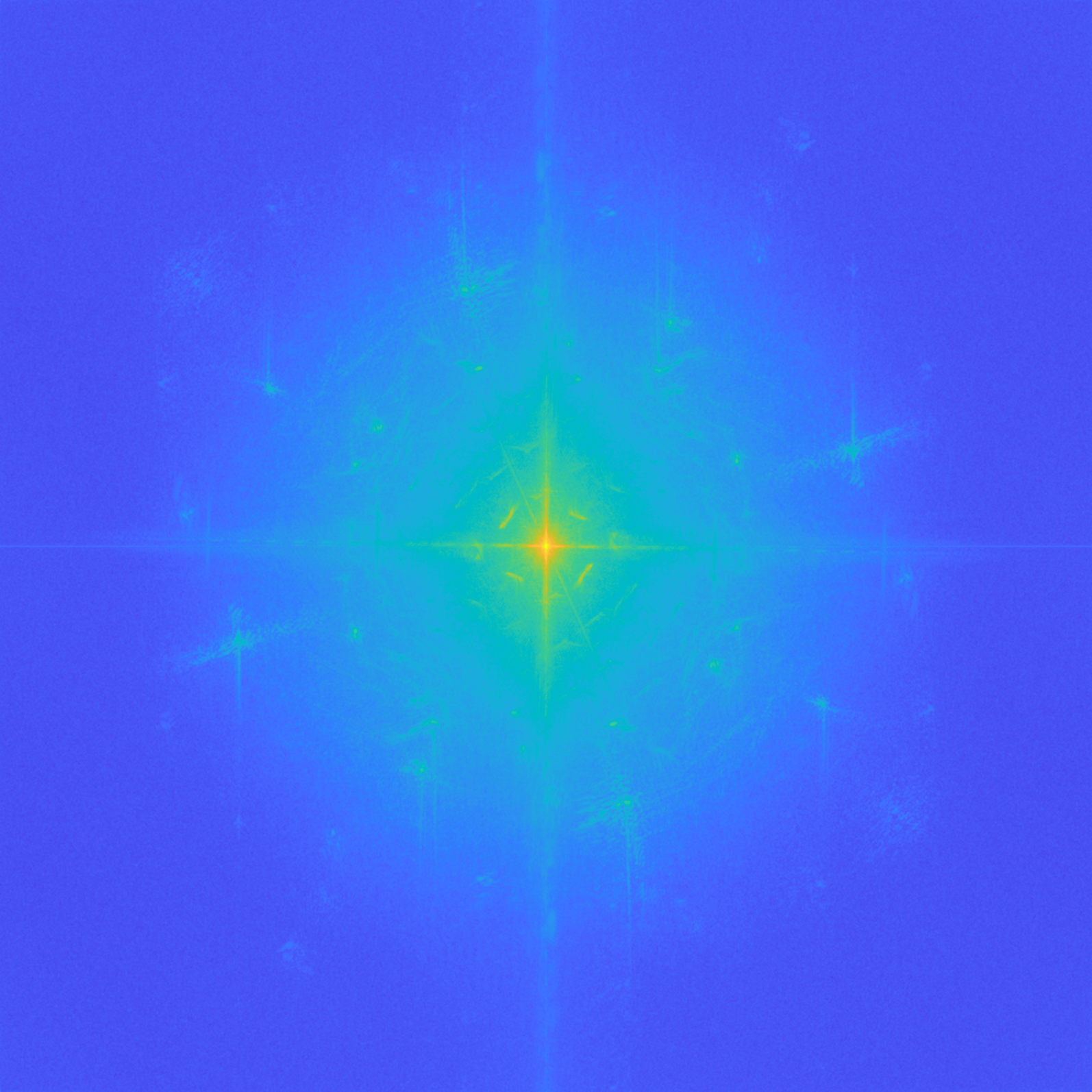}&
\includegraphics[width=.16\textwidth,height=.16\textwidth]{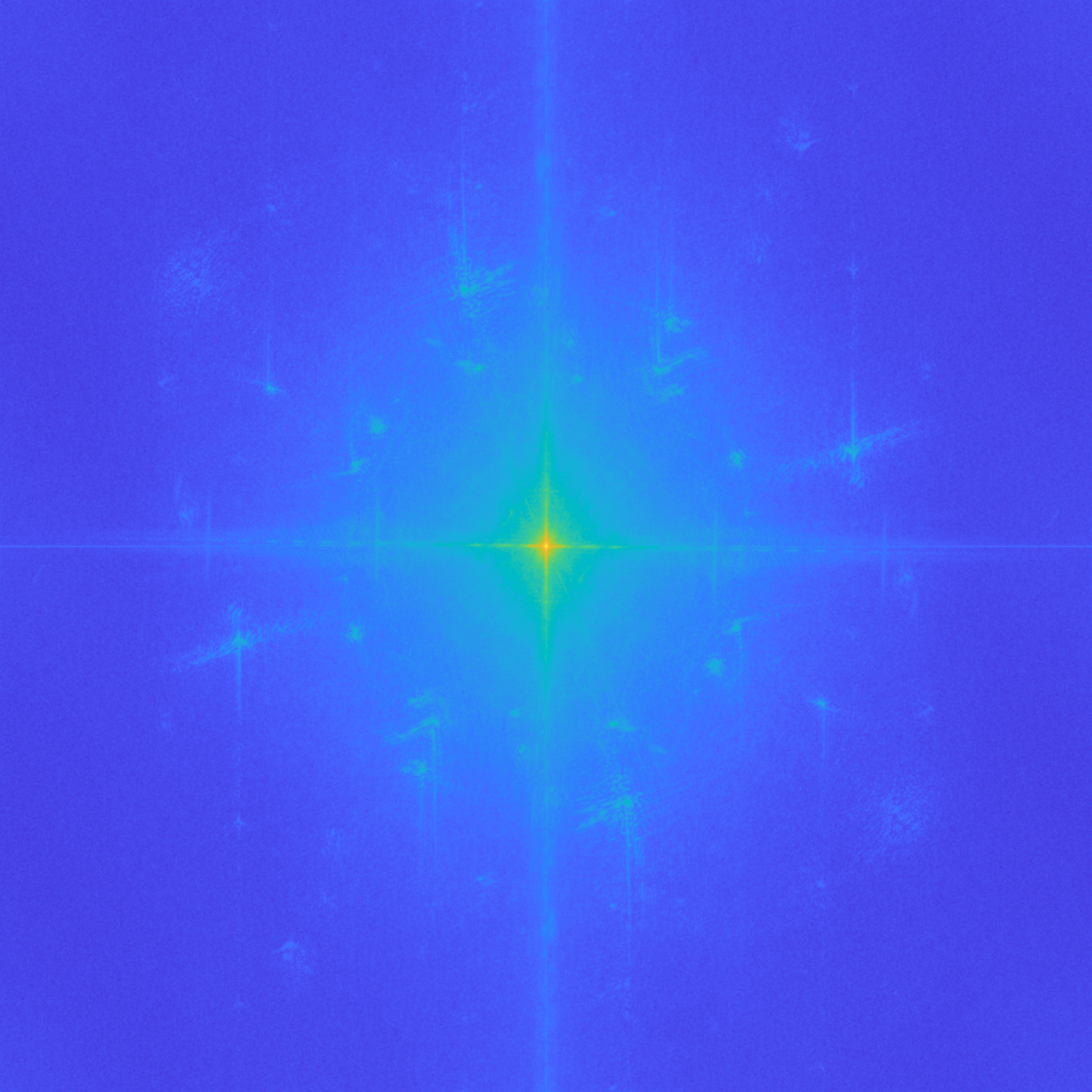}&
\includegraphics[width=.16\textwidth,height=.16\textwidth]{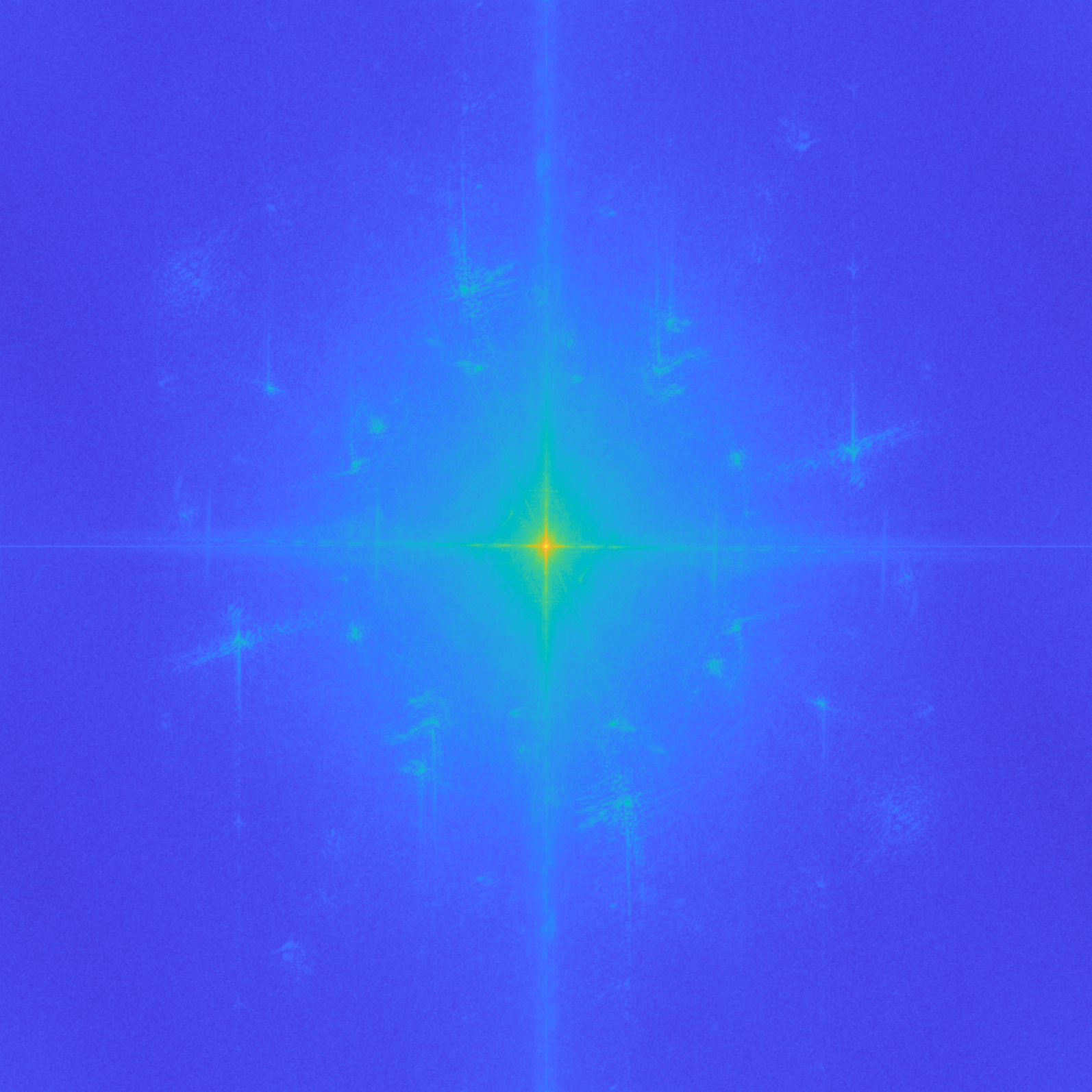}&&\\
(a)&(b)&(c)&&\\
\includegraphics[width=.16\textwidth,height=.16\textwidth]{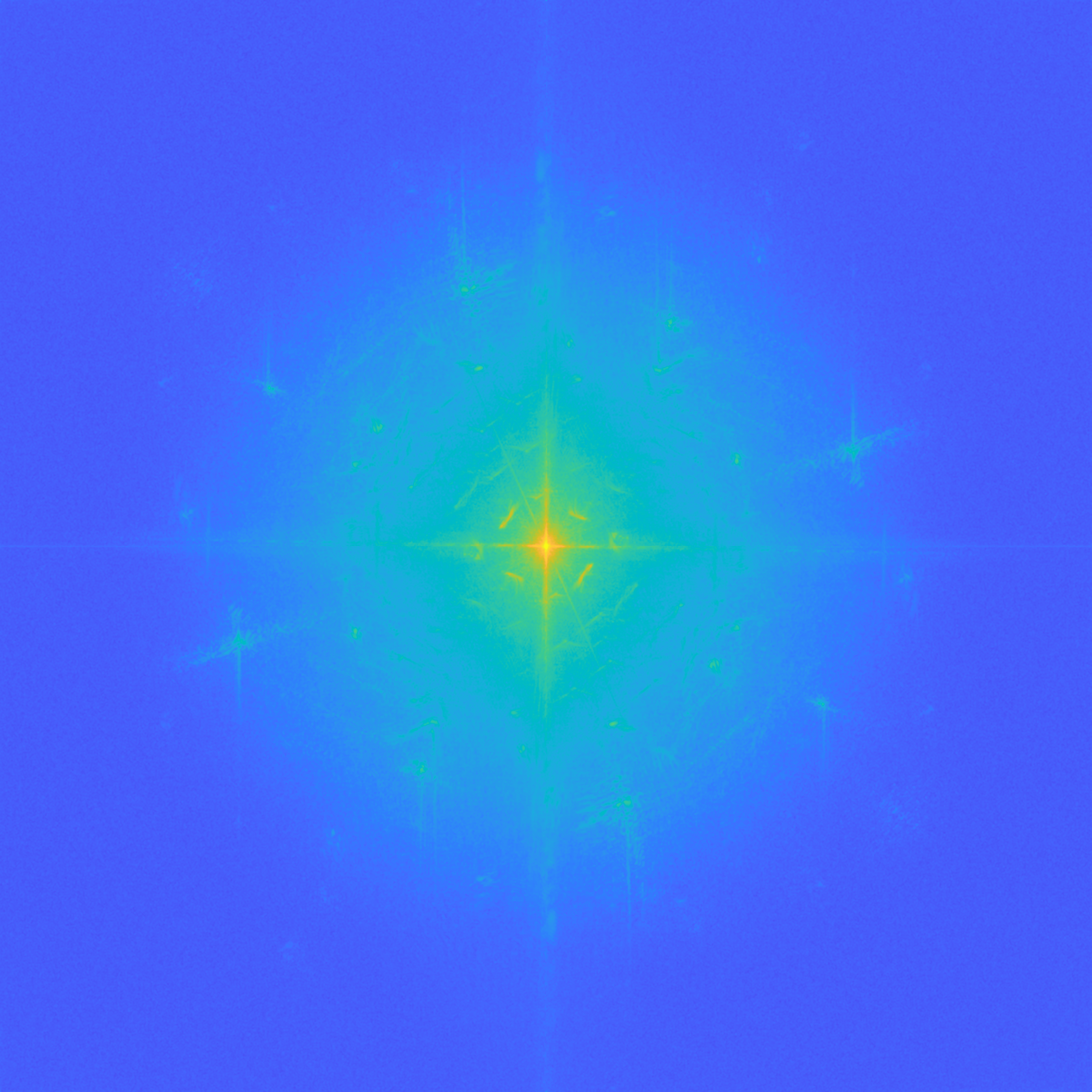}&
\includegraphics[width=.16\textwidth,height=.16\textwidth]{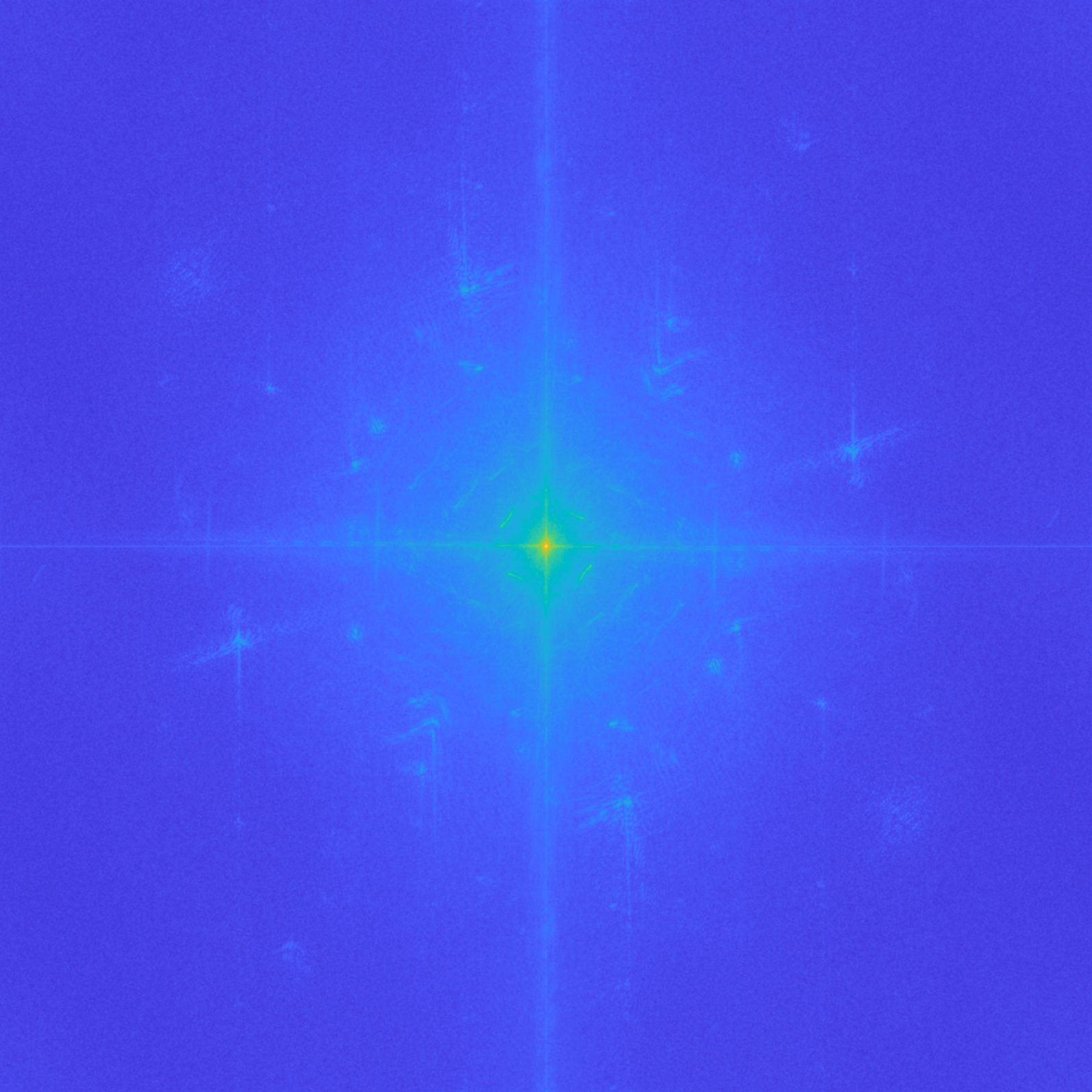}&
\includegraphics[width=.16\textwidth,height=.16\textwidth]{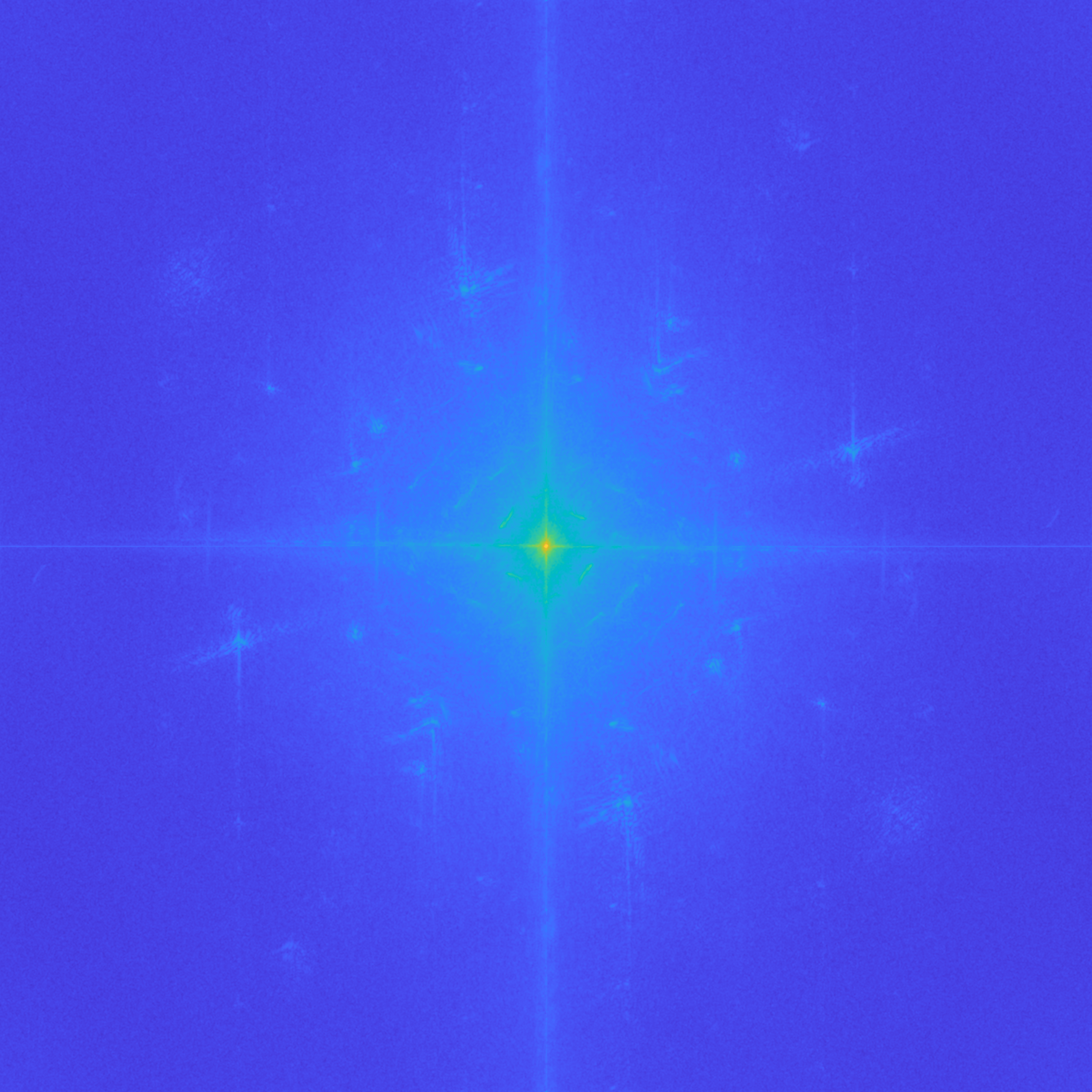}&
\multirow[t]{3}{*}{\includegraphics[width=.24\textwidth,height=.24\textwidth]{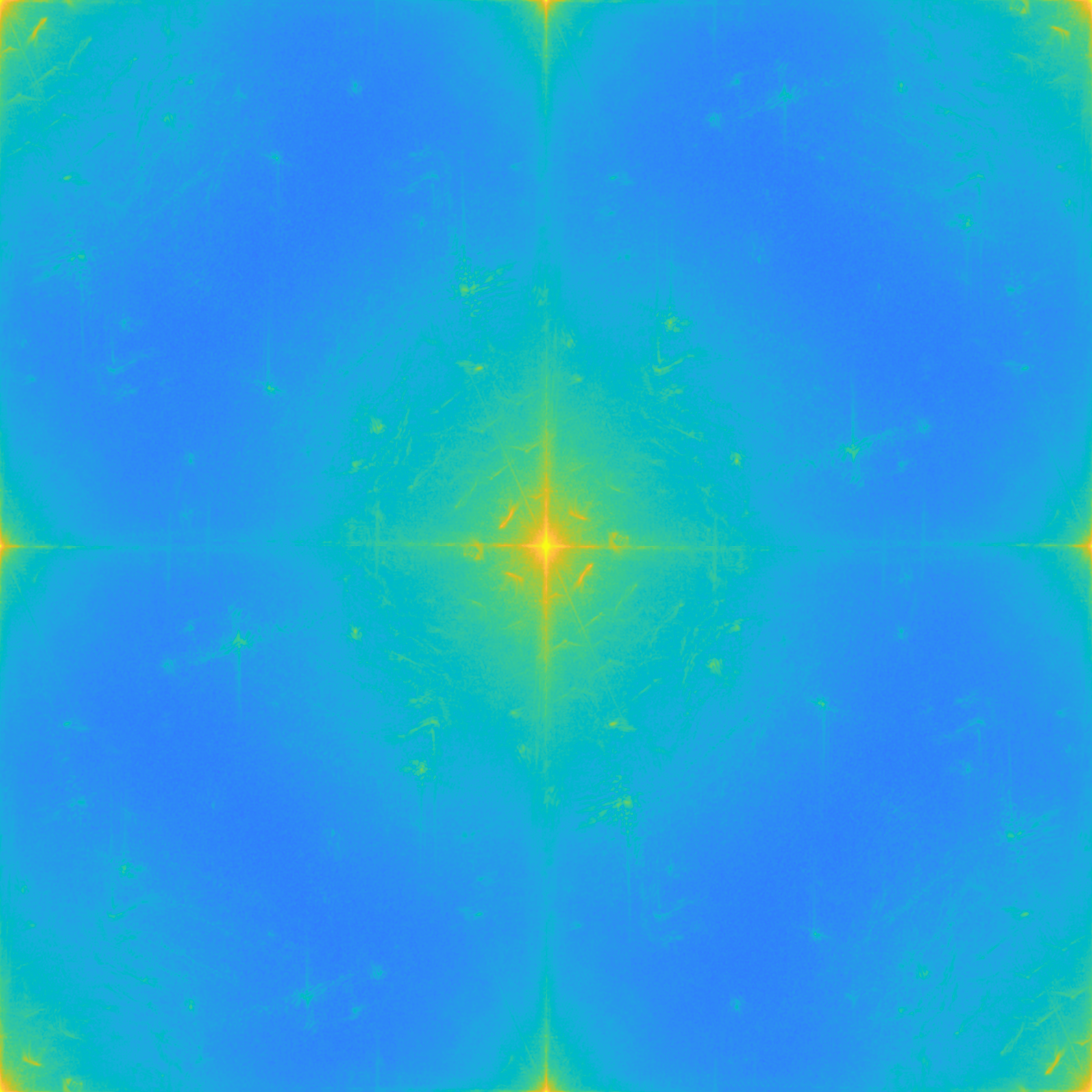}}&
\multirow[t]{3}{*}{\includegraphics[width=.24\textwidth,height=.24\textwidth]{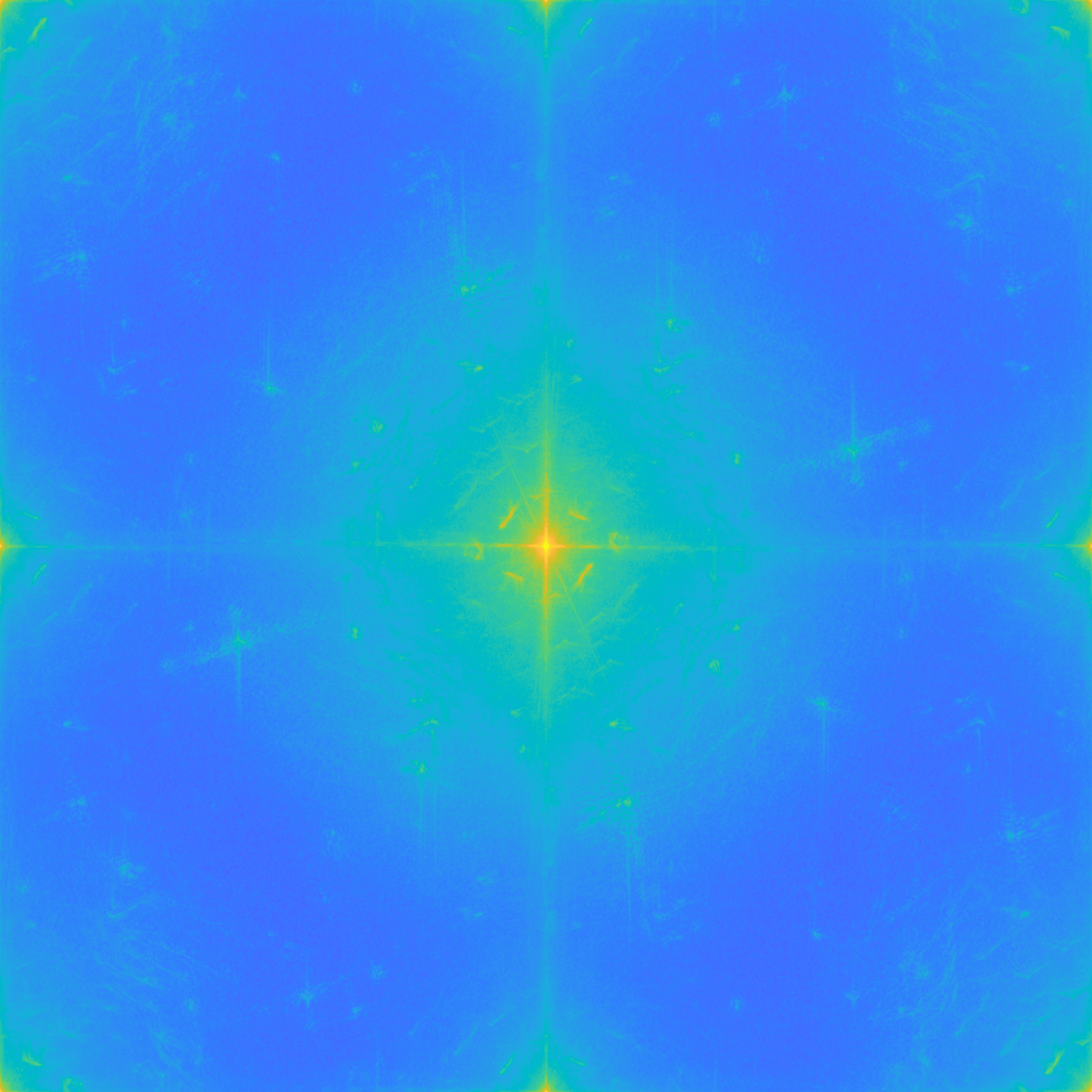}}\\
(d)&(e)&(f)&(g)&(h)\\
\end{tabular}
\caption{Fourier analysis of color image signals with and without white balance. DC component is at the center. Log Fourier magnitude of (a) Luminance $y_{\mu}$, (b) chrominance $y_{\gamma}$, (c) chrominance $y_{\beta}$, (d) white balanced luminance $z_{\mu}$, (e) white balanced chrominance $z_{\gamma}$, (f) white balanced chrominance $z_{\beta}$, (g) raw sensor data $x$, and (h) white-balanced raw sensor data $w$. DC is at the middle. White balance has the effect of reducing the spatial frequency support of the chrominances $z_{\gamma}$ and $z_{\beta}$. As a result, the risk of aliasing in (h) due to CFA sampling---which spatially modulates the chrominance signals---reduces significantly (compared to (g)). Computed from raw sensor data from Sony $\alpha$7R4. Additionally, pixel shift was used to bypass demosaicking procedure to generate images in (a)-(f).}
\label{fig:fft}
\end{figure*}

Recall \eqref{eq:spectralResponse}. In conventional cameras, there are three major reasons why the signal strengths of $y_r(i,j)$, $y_g(i,j)$, and $y_b(i,j)$ are very unbalanced. First, the quantum efficiency of CMOS and CCD sensors is typically poor in the blue color region, while it peaks in the green color region. Thus the signal in $y_b(i,j)$ is far weaker than $y_g(i,j)$ and $y_r(i,j)$. Second, the spectral transmittance of the color filters themselves might differ, meaning some filters may be more opaque than another. Lastly, the illumination spectra $L(\lambda)$ of the most natural and artificial light sources are not constant. The end result is that the raw sensor data appears to be shifted toward one color channel. the bandwidth of the chrominance signals. As shown by \fref{fig:fft}(g), increased spatial bandwidth of $y_{\gamma}(i,j)$ and $y_{\beta}(i,j)$ would worsen the risk of aliasing, making it difficult to decouple from $y_{\mu}(i,j)$.  

Let $\left[ \ell_r, \ell_g, \ell_b\right]$ denote the observed illumination color:
\begin{align} \label{eq:illuminant}
\begin{split}
\begin{bmatrix}
\ell_r\\ \ell_g\\ \ell_b
\end{bmatrix}
 =& \int \begin{bmatrix}\bar{r}(\lambda)\\\bar{g}(\lambda)\\\bar{b}(\lambda)\end{bmatrix}
 L(\lambda) d\lambda.
\end{split}
\end{align}
In post-processing, a task called ``white balance'' aims to restore the balance of the color channels. Inspired in part by the notion of ``color constancy'' in human visual system that makes color perception (nearly) invariant to illumination color~\cite{land1983recent, Finlayson93, Gijsenij11, Rang14}, the white balance  is usually carried out by normalizing by the illumination color:
\begin{align} \label{eq:whiteBalance} 
\vec{z}(i,j)=\begin{bmatrix} z_r(i,j) \\ z_g(i,j) \\ z_b(i,j) \end{bmatrix} = 
\begin{bmatrix} 1/\ell_r&0&0\\ 0&1/\ell_g&0\\ 0&0&1/\ell_b \end{bmatrix} 
\begin{bmatrix} y_r(i,j) \\ y_g(i,j) \\ y_b(i,j) \end{bmatrix}.
\end{align}

\subsection{Impact of White Balance on CFA Compression}
\label{sec:motivation}

The white balancing operation can help improve the overall coding efficiency. To understand why this is the case, consider a well-established and well-accepted model in demosaicking literature that the spatial highpass components of $z_r(i,j)$, $z_g(i,j)$, and $z_b(i,j)$ are similar. That is:
\begin{align}\label{eq:lp_hp}
\begin{split}
    \vec{z}(i,j) & =\begin{bmatrix}
    z_r^{\scriptscriptstyle LP}(i,j)+z_r^{\scriptscriptstyle HP}(i,j)\\
    z_g^{\scriptscriptstyle LP}(i,j)+z_g^{\scriptscriptstyle HP}(i,j)\\
    z_b^{\scriptscriptstyle LP}(i,j)+z_b^{\scriptscriptstyle HP}(i,j)
    \end{bmatrix} \vspace{0.2cm}\\ 
     & \approx\begin{bmatrix} 
     z_r^{\scriptscriptstyle LP}(i,j)\\
     z_g^{\scriptscriptstyle LP}(i,j)\\
     z_b^{\scriptscriptstyle LP}(i,j)
    \end{bmatrix}
    +z^{\scriptscriptstyle HP}(i,j)\begin{bmatrix}
    1\\1\\1
    \end{bmatrix},
\end{split}
\end{align}
where superscripts $LP$ and $HP$ refer to lowpass and highass components, and the $z^{\scriptscriptstyle HP}(i,j)$ is the common highpass components shared by all color channels.

Substituting \eqref{eq:whiteBalance} to \eqref{eq:lp_hp} and \eqref{eq:opponentcolor}, we have
\begin{align}\label{eq:y_lp_hp}
\vec{y}
=&    
\begin{bmatrix} 
    \ell_r \cdot z_r^{\scriptscriptstyle LP}\\
    \ell_g \cdot z_g^{\scriptscriptstyle LP}\\
    \ell_b \cdot  z_b^{\scriptscriptstyle LP}
    \end{bmatrix}
    +z^{\scriptscriptstyle HP}\begin{bmatrix}
    \ell_r\\\ell_g\\\ell_b
    \end{bmatrix}\\
    y_{\gamma}  =& \frac{\ell_r\cdot z_r^{\scriptscriptstyle LP}-\ell_b\cdot z_b^{\scriptscriptstyle LP}}{4}+\frac{\ell_r-\ell_b}{4} z^{\scriptscriptstyle HP}\notag \\
    y_{\beta} = &\frac{\ell_r\cdot z_r^{\scriptscriptstyle LP}-2\cdot \ell_g\cdot z_g^{\scriptscriptstyle LP}+\ell_b\cdot z_b^{\scriptscriptstyle LP}}{4}+\frac{\ell_r-2\cdot \ell_g+\ell_b}{4}z_b^{\scriptscriptstyle HP}.\notag
\end{align}
The analysis above implies that the degree of color imbalance $\ell_r-\ell_b$ and $\ell_r-2\cdot \ell_g+\ell_b$ determines how much the spatial highpass signal $z^{\scriptscriptstyle HP}(i,j)$ contributes to the chrominance signals $y_{\gamma}(i,j)$ and $y_{\beta}(i,j)$. That is, the chrominance signals are expected to be spatially broadband, unless $\ell_r=\ell_g=\ell_b$. Thus, the CFA sampled data $x(i,j)$ is unfavorable to demosaicking (and potentially harmful to CFA compression, as we demonstrate below).

Recall \eqref{eq:spectralResponse}. Suppose white balance is applied to CFA sampled data by normalizing red, green, and blue filtered values by $\ell_r$, $\ell_g$, and $\ell_b$, respectively. That is:
\begin{align}\label{eq:cfa_wb}
\begin{split}
    w(i,j)=&
    \begin{cases}
    x(i,j)/\ell_r&\text{if $\vec{c}(i,j)=[1,0,0]^T$}\\
    x(i,j)/\ell_g&\text{if $\vec{c}(i,j)=[0,1,0]^T$}\\
    x(i,j)/\ell_b&\text{if $\vec{c}(i,j)=[0,0,1]^T$}
    \end{cases}\\
    =&\vec{c}(i,j)^T
    \begin{bmatrix}
    y_r(i,j)/\ell_r\\
    y_g(i,j)/\ell_g\\
    y_b(i,j)/\ell_b
    \end{bmatrix}\\
    =&\vec{c}(i,j)^T\vec{z}(i,j).
\end{split}
\end{align}
That is, white balanced CFA data $w(i,j)$ results in CFA sampled version of the white balanced color image $\vec{z}(i,j)$. Thus substituting the signal model \eqref{eq:lp_hp} into the luminance-chrominance decomposition of CFA sampled data in \eqref{eq:opponentcolor} yields the following:
\begin{align}\label{eq:lp}
\begin{split}
    z_{\gamma} & = \frac{z_r^{\scriptscriptstyle LP}-z_b^{\scriptscriptstyle LP}}{4}+\frac{z^{\scriptscriptstyle HP}-z^{\scriptscriptstyle HP}}{4}=\frac{z_r^{\scriptscriptstyle LP}-z_b^{\scriptscriptstyle LP}}{4}\\
    z_{\beta} & = \frac{z_r^{\scriptscriptstyle LP}-2z_g^{\scriptscriptstyle LP}+z_b^{\scriptscriptstyle LP}}{4}+\frac{z^{\scriptscriptstyle HP}-2z^{\scriptscriptstyle HP}+z^{\scriptscriptstyle HP}}{4} \\ 
    & = \frac{z_r^{\scriptscriptstyle LP}-2z_g^{\scriptscriptstyle LP}+z_b^{\scriptscriptstyle LP}}{4}.
\end{split}
\end{align}
We may conclude that the white balanced chrominance signals $z_{\gamma}(i,j)$ and $z_{\beta}(i,j)$ are lowpass signals, thanks to the cancellation of highpass components. 

\fref{fig:fft}(e,f,h) shows empirical evidence confirming that white balance reduces the spatial bandwidth of chrominance signals. Specifically, the Fourier supports of $z_{\gamma}$ and $z_{\beta}$ are significantly reduced compared to that of $y_{\gamma}$ and $y_{\beta}$, respectively. As a result, the luminance and chrominance components are more decoupled in the Fourier magnitudes of CFA sampled signal $w$ than the Fourier magnitudes of $x$. Indeed, white balanced raw sensor data in \fref{fig:bayerPattern}(c) is smooth overall, thanks to the fact that modulated chrominance components have very little spatial structure to add (i.e.~spatially lowpass). As a result, pre-demosaicking white balance helps the demosaicking performance.

The reduced spatial bandwidth of the white balanced chrominance images  $z_{\gamma}$ and $z_{\beta}$ presents a significant opportunity for CFA compression. There are at least two categories of CFA compression methods that are likely to benefit from encoding CFA sampled image $w(i,j)$ instead of raw sensor data $x(i,j)$. First, as evidenced by \fref{fig:fft}, Fourier support of $w$ is more sparse than that of $x$. Thus compression techniques that are designed to exploit spatial frequency decompositions would enjoy higher coding efficiencies. This includes Mallat wavelet method~\cite{Mallat98} as well as the CAMRA-A~\cite{lee2018camera} and CAMRA-S~\cite{lee2020shift} methods that further decorrelate the wavelet decomposition.

The second category of CFA compression methods that benefit from white balance is those that explicitly perform decorrelation across color components. These decorrelation tasks have effects similar to \eqref{eq:lp} where the highpass signals cancel, thereby achieving higher compression efficiency. Methods falling in this category include MSST~\cite{Malvar12}, CAMRA-A ~\cite{lee2018camera}, and CAMRA-S~\cite{lee2020shift}.

\section{Proposed: Lossless White Balance}
\label{sec:proposed}

The main conclusion we draw from the CFA sampling analysis in Section \ref{sec:cfaSampling} is that the color imbalance problem negatively affects raw sensor data compression. Although the white balance operation in \eqref{eq:cfa_wb} as a preprocessing to CFA image/video compression would help restore the cross-color correlation and improve coding efficiency, non-integer implementation of white balance would incur penalty through quantization, making it impossible to encode losslessly.

In this section, we develop a novel lossless white balance technique using lifting. Specifically, we develop a lifting scheme for scalar multiplication in Section \ref{sec:scalarLifting}. In Section \ref{sec:wb_lifting}, we combine several scalar multiplication lifting schemes to derive an exactly reversible white balance procedure.

\subsection{Lifting For Lossless Scalar Multiplication} \label{sec:scalarLifting}
% Fig. 3
\begin{figure*}[t]
\begin{minipage}{0.49\linewidth}
    \centering
    \centerline{\includegraphics[scale=0.55]{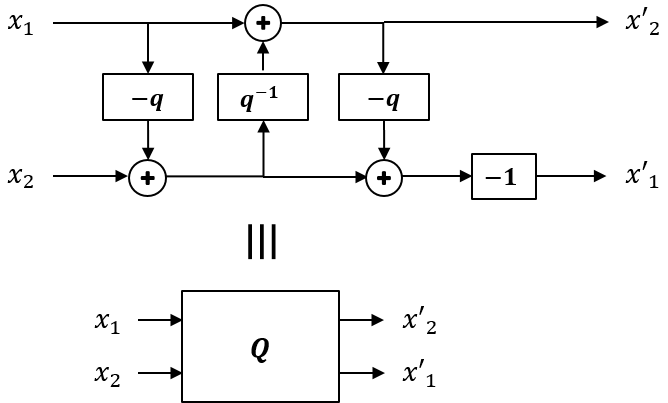}}
    \centerline{\footnotesize{(a)}}
\end{minipage}
\begin{minipage}{0.49\linewidth}
    \centering
    \centerline{\includegraphics[scale=0.55]{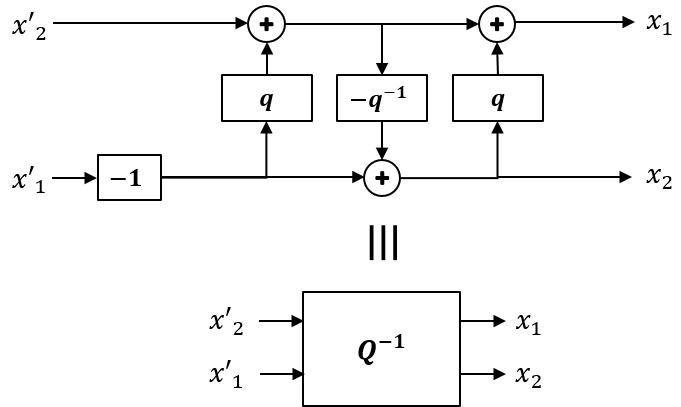}}
    \centerline{\footnotesize{(b)}}
\end{minipage}
\caption{Scalar multiplication lifting scheme. (a) Forward lifting steps. (b) Inverse lifting steps. The prediction and update steps use scalar multiplication only.}
\label{fig:structureQ}
\end{figure*}
Borrowing the structure of \fref{fig:liftingDWT}, the reversible lifting structure can have multiple predict and update steps. Consider a lifting configuration as shown in \fref{fig:structureQ}(a), which performs the following set of matrix operations:
\begin{align}\label{eq:structureQ}
     \begin{bmatrix}  x_2^\prime\\x_1^\prime \end{bmatrix} &  = 
\begin{bmatrix}  1&0 \\0&-1 \end{bmatrix}
\begin{bmatrix}  1&0 \\-q&1 \end{bmatrix}
\begin{bmatrix}  1&1/q \\0&1 \end{bmatrix}
\begin{bmatrix}  1&0 \\-q&1 \end{bmatrix}
\begin{bmatrix}  x_1\\x_2 \end{bmatrix}\\
=&
\underbrace{
\begin{bmatrix}0&1/q\\ q&0\end{bmatrix}}_{\vec{Q}}
\begin{bmatrix} x_1\\x_2 \end{bmatrix}.\label{eq:Q}
\end{align}
In other words, we accomplish a scalar multiplication of $\left[x_e\:\:x_o\right]^T$ by this series of prediction and update steps. The advantage to the expanded form in \eqref{eq:structureQ} over the direct application of matrix $\vec{Q}$ in \eqref{eq:Q} is that the scaling is done losslessly using lifting scheme with appropriate rounding operations. Specifically, procedure in \eqref{eq:structureQ} can be implemented by the lifting scheme shown in Figure \ref{fig:structureQ}(a), whose steps are detailed in Algorithm \ref{alg:scalar_forward}.

\begin{algorithm}[t]
\caption{Forward scalar lifting structure}
\label{alg:scalar_forward}
\begin{algorithmic}
\STATE $x_2\gets  x_2-\lfloor q\cdot x_1\rfloor$
\STATE $x_1\gets  x_1+\lfloor x_2/q\rfloor$
\STATE $x_2\gets x_2- \lfloor  q\cdot x_1\rfloor$
\STATE $(x_1',x_2')\gets (-x_2,x_1)$
\end{algorithmic}
\end{algorithm}

As before, the lifting structure in \eqref{eq:structureQ} can be inverted by reversing each predict/update operations as described in \fref{fig:structureQ}(b):
\begin{align}\label{eq:structureQinverse}
\begin{split}
\begin{bmatrix}  x_1\\x_2 \end{bmatrix}
=&
\underbrace{
\begin{bmatrix}  1&q \\0&1 \end{bmatrix}
\begin{bmatrix}  1&0 \\-1/q&1 \end{bmatrix}
\begin{bmatrix}  1&q \\0&1 \end{bmatrix}
\begin{bmatrix}  1&0 \\0&-1 \end{bmatrix}}_{Q^{-1}}
\begin{bmatrix}  x_2^\prime\\x_1^\prime \end{bmatrix}.
\end{split}
\end{align}
The lossless implementation of \eqref{eq:structureQinverse} is illustrated in Figure \ref{fig:structureQ}(b), whose steps are detailed in Algorithm \ref{alg:scalar_inverse}. Contrast this to the direct application of matrix $\vec{Q}$ and $\vec{Q}^{-1}$ onto $\left[x_1\:\:x_2\right]^T$, which would result in loss incurred by precision rounding errors.

\begin{algorithm}[t]
\caption{Inverse scalar lifting structure}
\label{alg:scalar_inverse}
\begin{algorithmic}
\STATE $x_2'\gets  x_2'-\lfloor q\cdot x_1'\rfloor$
\STATE $x_1'\gets  -x_1'-\lfloor x_2'/q\rfloor$
\STATE $x_2'\gets  x_2'+\lfloor q\cdot x_1'\rfloor$
\STATE $(x_1,x_2)=(x_2',x_1')$
\end{algorithmic}
\end{algorithm}

It is important to point out that the scalar lifting scheme in \eqref{eq:structureQ} on its own has no practical impact on the coding efficiency, despite the changes in dynamic range due to multiplication/division by a scalar value $q\in\mathbb{R}$. To understand this, we use differential entropy and continuous variables as a proxy for discrete entropy for discrete symbols $(x_1,x_2)$, respectively. Thus the number of bits required for encoding $(x_1,x_2)$ is:
\begin{align}
    &h(x_1)+h(x_2)=\\
    &-\int f_1(x_1)\log_2 f_1(x_1) dx_1-\int f_2(x_2)\log_2 f_2(x_2) dx_2.\notag
\end{align}
where $f_1,f_2:\mathbb{R}\to\mathbb{R}_+$ are the probability density functions of $x_1$ and $x_2$, respectively. The probability density functions of $x_1'$ and $x_2'$ are:
\begin{align}
\begin{split}
    f_1'(x_1')=&\frac{f_1(x_1'/q)}{q}\\
    f_2'(x_2')=&q\cdot f_2(q\cdot x_2')
\end{split}
\end{align}
We compare $h(x_1,x_2)$ to the differential entropy of $(x_1',x_2')$:

\begin{align}\label{eq:entropy}
\begin{array}{l}
h(x_1')+h(x_2') \vspace{0.3cm}\\
=-\int f_1'(x_1')\log_2 f_1'(x_1') d x_1-\int f_2'(x_2')\log_2 f_2'(x_2')d x_2'\vspace{0.3cm}\\
=-\int f_1(x_1)\log_2 \frac{f_1(x_1)}{q} d x_1-\int f_2(x_2)\log_2 (q\cdot f_2(x_2)) d x_2 \vspace{0.3cm}\\
=(h(x_1)+\log_2 q)+(h(x_2)-\log_2 q)=h(x_1)+h(x_2).
\end{array}
\end{align}

In practice, discrete entropy and the rounding effects of the lifting scheme may alter the encoding efficiency by small amounts. But by-and-large, we conclude from \eqref{eq:entropy} that there is no advantage to coding $(x_1',x_2')$ over coding $(x_1,x_2)$ in absence of other techniques that take advantage of scalar lifting schemes. Indeed, this is the outcome we see in the compression performance of ``Demux'' in Table \ref{tab:bitrates} (See Section \ref{sec:results}).

\subsection{Lifting For Lossless White Balance}
\label{sec:wb_lifting}

% Fig.4
\begin{figure*}[t]
\begin{minipage}{0.49\linewidth}
    \centering
    \centerline{\includegraphics[scale=0.55]{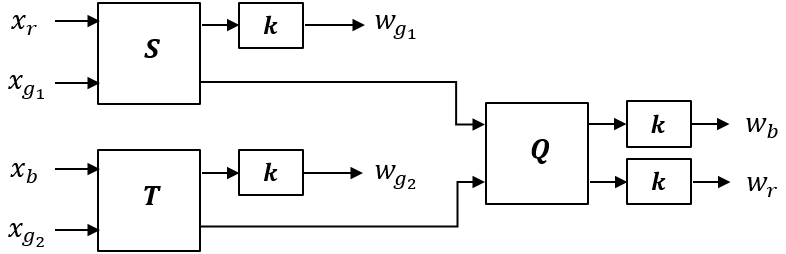}}
    \centerline{\footnotesize{(a)}}
\end{minipage}
\begin{minipage}{0.49\linewidth}
    \centering
    \centerline{\includegraphics[scale=0.55]{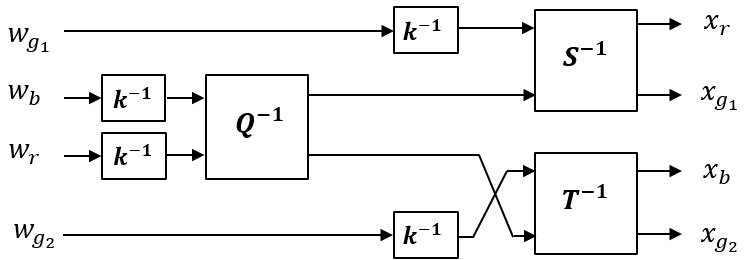}}
    \centerline{\footnotesize{(b)}}
\end{minipage}
\caption{Proposed lossless white balance lifting scheme in ``pyramid'' structure. (a) Forward lifting steps. (b) Inverse lifting steps. The blocks $Q$, $S$, and $T$ refer to the scalar multiplication lifting scheme in \fref{fig:structureQ}(a). Their inverses use the lifting scheme in \fref{fig:structureQ}(b).}
\label{fig:losslessWhiteBalance}
\end{figure*}

The transforms described in \eref{eq:whiteBalance} and \eqref{eq:cfa_wb} are irreversible in a lossless manner. We instead leverage the integer-based multiplication using the lifting structure described in \fref{fig:structureQ}. Given the Bayer patterned sensor array in \fref{fig:bayerPattern}, the color channels split into $x_r, x_{g_1}, x_{g_2}, x_b$ corresponding to their locations:
\begin{align}
    \begin{split}
        x_r(i,j)=&x(2i,2j)\\
        x_{g_1}(i,j)=&x(2i,2j+1)\\
        x_{g_2}(i,j)=&x(2i+1,2j)\\
        x_b(i,j)=&x(2i+1,2j+1).
    \end{split}
\end{align}
Define $w_r,w_{g_1},w_{g_2},w_b$ in a similar way. Recalling \eref{eq:cfa_wb}, the desired white balance operation is follows: 
\begin{align}\label{eq:losslessWhiteBalance}
\begin{bmatrix} w_r(i,j)\\w_{g_1}(i,j)\\w_{g_2}(i,j)\\w_b(i,j) \end{bmatrix}  = 
\begin{bmatrix} \bar{\ell}/\ell_r&0&0&0\\ 0&\bar{\ell}/\ell_{g_1}&0&0\\ 0&0&\bar{\ell}/\ell_{g_2}&0\\ 0&0&0&\bar{\ell}/\ell_b\end{bmatrix}
\begin{bmatrix} x_r(i,j)\\x_{g_1}(i,j)\\x_{g_2}(i,j)\\x_b(i,j) \end{bmatrix},
\end{align}
where the observed illumination color is $\left[ \ell_r, \ell_{g_1}, \ell_{g_2},\ell_b\right]$, and their geometric mean
\begin{align}
    \bar{\ell}=(\ell_r\ell_{g_1}\ell_{g_2}\ell_b)^{\frac{1}{4}}
\end{align}
keeps the dynamic range of
$w(i,j)$ similar to $x(i,j)$. Here, we have allowed the possibility that $x_{g_1}$ and $x_{g_2}$ have different white balance correction values $\bar{\ell}/\ell_{g_1}$ and $\bar{\ell}/\ell_{g_2}$, respectively. This is sometimes necessary when there is a crosstalk or misalignment between the CFA and the pixel sensor array
~\cite{Hirakawa08Crosstalk}. But the theory and the method we present below is equally valid when $\ell_{g_1}=\ell_{g_2}$.

We propose to implement a lossless white balance operation by carrying out \eqref{eq:losslessWhiteBalance} using a series of lifting structures defined in \fref{fig:structureQ}. Consider a lifting structure in  \fref{fig:losslessWhiteBalance}(a). Mathematically, it is equivalent to:
\begin{align}\label{eq:losslessWhiteBalanceQ}\begin{array}{rl}
\begin{bmatrix} w_r(i,j)\\w_{g_1}(i,j)\\w_{g_2}(i,j)\\w_b(i,j) \end{bmatrix}  & 
= \vec{K}\vec{Q}\vec{T}\vec{S}
\begin{bmatrix}x_r(i,j)\\x_{g_1}(i,j)\\x_{g_2}(i,j)\\x_b(i,j)\end{bmatrix} \vspace{0.2cm}\\
& = \begin{bmatrix} 
ksq&0&0&0\\
0&k/s&0&0\\
0&0&k/t&0\\
0&0&0&kt/q
\end{bmatrix}
\begin{bmatrix}x_r(i,j)\\x_{g_1}(i,j)\\x_{g_2}(i,j)\\x_b(i,j) \end{bmatrix},
\end{array}\end{align}
where
\begin{align}\begin{array}{l}
\vec{S}=\begin{bmatrix} 0&1/s&0&0\\s&0&0&0\\0&0&1&0\\0&0&0&1\end{bmatrix},
\vec{T} = \begin{bmatrix} 1&0&0&0\\0&1&0&0\\0&0&0&t\\0&0&1/t&0\end{bmatrix}\vspace{0.2cm}\\
\vec{Q} = \begin{bmatrix} 
1&0&0&0\\
0&0&1/q&0\\
0&q&0&0\\
0&0&0&1
\end{bmatrix}, \vec{K} = \begin{bmatrix} 0&0&k&0\\k&0&0&0\\0&0&0&k\\0&k&0&0\end{bmatrix} \end{array}
\end{align}
are the pairwise lifting operations. The lifting coefficients $(s,t,q,k)$ must be chosen appropriately in order for the lifting structure in \fref{fig:losslessWhiteBalance}(a) and \eref{eq:losslessWhiteBalanceQ} to match the behavior of \eref{eq:losslessWhiteBalance}. Towards this goal, consider representing
the white balance correction values  in \eqref{eq:losslessWhiteBalance} using logarithm functions:
\begin{align}\label{eq:logwhitebalance}
    \begin{bmatrix}
    \ln (\bar{\ell}/\ell_r)\\ 
    \ln (\bar{\ell}/\ell_{g_1})\\
    \ln (\bar{\ell}/\ell_{g_2})\\
    \ln (\bar{\ell}/\ell_b)
    \end{bmatrix}
=&
\begin{bmatrix}
-3/4&1/4&1/4&1/4\\
1/4&-3/4&1/4&1/4\\
1/4&1/4&-3/4&1/4\\
1/4&1/4&1/4&-3/4\\
\end{bmatrix}
\begin{bmatrix} \ln\ell_r \\ \ln\ell_{g_1}\\\ln\ell_{g_2}\\\ln\ell_{b}\end{bmatrix}.
\end{align}
We may also represent the  lifting coefficients in \eqref{eq:losslessWhiteBalanceQ}
in the log domain:
\begin{align}\label{eq:loglifting}
    \begin{bmatrix}
    \ln (\bar{\ell}/\ell_r)\\ 
    \ln (\bar{\ell}/\ell_{g_1})\\
    \ln (\bar{\ell}/\ell_{g_2})\\
    \ln (\bar{\ell}/\ell_b)
    \end{bmatrix}
    =&
\begin{bmatrix} 1&0&1&1\\-1&0&0&1\\0&-1&0&1\\0&1&-1&1\end{bmatrix}
\begin{bmatrix} \ln s\\\ln t\\\ln q\\\ln k\end{bmatrix}.
\end{align}
Combining \eqref{eq:logwhitebalance} and \eqref{eq:loglifting} we solve a system of linear equations, as follows:
\begin{align}\label{eq:equivWhiteBalance}
% \begin{split}
% &\begin{bmatrix}
% -3/4&1/4&1/4&1/4\\
% 1/4&-3/4&1/4&1/4\\
% 1/4&1/4&-3/4&1/4\\
% 1/4&1/4&1/4&-3/4\\
% \end{bmatrix}
% \begin{bmatrix} \ln\ell_r \\ \ln\ell_{g_1}\\\ln\ell_{g_2}\\\ln\ell_{b}\end{bmatrix} \\
% &= 
% \begin{bmatrix} 1&0&1&1\\-1&0&0&1\\0&-1&0&1\\0&1&-1&1\end{bmatrix}
% \begin{bmatrix} \ln s\\\ln t\\\ln q\\\ln k\end{bmatrix}.
%\end{split}
\begin{bmatrix} \ln s\\\ln t\\\ln q\\\ln k\end{bmatrix}
= \begin{bmatrix}
-1&3&-1&-1\\
-1&-1&3&-1\\
-2&-2&2&2\\
0&0&0&0
\end{bmatrix}
\begin{bmatrix} \ln\ell_r \\ \ln\ell_{g_1}\\\ln\ell_{g_2}\\\ln\ell_{b}\end{bmatrix},
\end{align}
% Inverting this linear system, we recover the appropriate lifting coefficients as:
% \begin{align}
% \begin{bmatrix} \ln s\\\ln t\\\ln q\\\ln k\end{bmatrix}
% = \begin{bmatrix}
% -1&3&-1&-1\\
% -1&-1&3&-1\\
% -2&-2&2&2\\
% 0&0&0&0
% \end{bmatrix}
% \begin{bmatrix} \ln\ell_r \\ \ln\ell_{g_1}\\\ln\ell_{g_2}\\\ln\ell_{b}\end{bmatrix},
% \end{align}
or equivalently,
\begin{align}\label{eq:coefficients}
\begin{array}{ll}
s = \left( \frac{\ell_{g_1}^3}{\ell_{r}\ell_{g_2}\ell_{b}}\right)^{1/4}, & t = \left(\frac{\ell_{g_2}^3}{\ell_{r}\ell_{g_1}\ell_{b}} \right)^{1/4}, \vspace{0.2cm}\\
q = \left( \frac{\ell_b\ell_{g_2}}{\ell_r\ell_{g_1}}\right)^{1/2}, & k = 1.
\end{array}\end{align}
% where
% \begin{align}
% \begin{bmatrix} \ln s\\\ln t\\\ln q\\\ln k\end{bmatrix} =  \begin{bmatrix} -1&0&-1&-1\\1&0&0&-1\\0&1&0&-1\\0&-1&1&-1\end{bmatrix}^{-1} \begin{bmatrix} \ln\ell_r \\ \ln\ell_{g_1}\\\ln\ell_{g_2}\\\ln\ell_{b}\end{bmatrix}
% \end{align}

The white balance procedure in \eref{eq:losslessWhiteBalanceQ} can be implemented with the lifting structure of \fref{fig:losslessWhiteBalance}(a) using the coefficients in \eqref{eq:coefficients}, steps of which are outlined in Algorithm \ref{alg:wb_forward}. This is perfectly invertible by another lifting structure in \fref{fig:losslessWhiteBalance}(b), carrying out the following operations:
\begin{align}\label{eq:inverse_lifting}
    \begin{bmatrix} x_r(i,j)\\x_{g_1}(i,j)\\x_{g_2}(i,j)\\x_b(i,j) \end{bmatrix}  =
    \vec{S}^{-1}\vec{T}^{-1}\vec{Q}^{-1}\vec{K}^{-1}
    \begin{bmatrix} w_r(i,j)\\w_{g_1}(i,j)\\w_{g_2}(i,j)\\w_b(i,j) \end{bmatrix}.
\end{align}
Exact reconstruction can be achieved when \eqref{eq:inverse_lifting} is implemented as shown in Algorithm \ref{alg:wb_inverse}. Obviously, lifting coefficients $s,t,q$ need to be stored or transmitted to the decoder as side information in order to invert the white balance procedure. But since they are global variables, the bitrate overhead is negligible.

\begin{algorithm}[t]
\begin{algorithmic}
\caption{Forward white balance lifting structure}
\label{alg:wb_forward}
\STATE $I\times J$ is the image size of $x$.
\FOR{$i=0$ to $I/2-1$}
\FOR{$j=0$ to $J/2-1$}
\STATE $x_r\gets x(2i,2j)$
\STATE $x_{g_1}\gets x(2i,2j+1)$
\STATE $x_{g_2}\gets x(2i+1,2j)$
\STATE $x_b\gets x(2i+1,2j+1)$
%%%%
\STATE $x_{g_1}\gets  x_{g_1}-\lfloor s\cdot x_r\rfloor$\\
\STATE $x_r\gets  x_r+\lfloor x_{g_1}/s\rfloor$\\
\STATE $x_{g_1}\gets x_{g_1}- \lfloor  s\cdot x_r\rfloor$\\
\STATE $(x_r,x_{g_1})\gets (-x_{g_1},x_r)$\\
%%%%
\STATE $x_{g_2}\gets  x_{g_2}-\lfloor t\cdot x_b\rfloor$\\
\STATE $x_b\gets  x_b+\lfloor x_{g_2}/t\rfloor$\\
\STATE $x_{g_2}\gets x_{g_2}- \lfloor  t\cdot x_b\rfloor$\\
\STATE $(x_b,x_{g_2})\gets (-x_{g_2},x_b)$\\
%%%%
\STATE $x_b\gets  x_b-\lfloor q\cdot x_r\rfloor$\\
\STATE $x_r\gets  x_r+\lfloor x_b/q\rfloor$\\
\STATE $x_b\gets x_b- \lfloor  q\cdot x_r\rfloor$\\
\STATE $(x_r,x_b)\gets (-x_b,x_r)$\\
%%%%
\STATE $w(2i,2j)\gets x_r$
\STATE $w(2i,2j+1)\gets x_{g_1}$
\STATE $w(2i+1,2j)\gets x_{g_2}$
\STATE $w(2i+1,2j+1)\gets x_b$
\ENDFOR
\ENDFOR
\end{algorithmic}
\end{algorithm}

\begin{algorithm}[t]
\begin{algorithmic}
\caption{Inverse white balance lifting structure}
\STATE $I\times J$ is the image size of $x$.
\label{alg:wb_inverse}
\FOR{$i=0$ to $I/2-1$}
\FOR{$j=0$ to $J/2-1$}
\STATE $w_r\gets w(2i,2j)$
\STATE $w_{g_1}\gets w(2i,2j+1)$
\STATE $w_{g_2}\gets w(2i+1,2j)$
\STATE $w_b\gets w(2i+1,2j+1)$
%%%%
\STATE $w_b\gets  w_b-\lfloor q\cdot w_r\rfloor$
\STATE $w_r\gets  -w_r-\lfloor w_b/q\rfloor$
\STATE $w_b\gets  w_b+\lfloor q\cdot w_r\rfloor$
\STATE $(w_r,w_b)=(w_b,w_r)$
%%%
\STATE $w_{g_2}\gets  w_{g_2}-\lfloor t\cdot w_b\rfloor$
\STATE $w_b\gets  -w_b-\lfloor w_{g_2}/t\rfloor$
\STATE $w_{g_2}\gets  w_{g_2}+\lfloor t\cdot w_b\rfloor$
\STATE $(w_b,w_{g_2})=(w_{g_2},w_b)$
%%%%
\STATE $w_{g_1}\gets  w_{g_1}-\lfloor q\cdot w_r\rfloor$
\STATE $w_r\gets  -w_r-\lfloor w_{g_1}/q\rfloor$
\STATE $w_{g_1}\gets  w_{g_1}+\lfloor q\cdot w_r\rfloor$
\STATE $(w_r,w_{g_1})=(w_{g_1},w_r)$
\STATE $x(2i,2j)\gets w_r$
\STATE $x(2i,2j+1)\gets w_{g_1}$
\STATE $x(2i+1,2j)\gets w_{g_2}$
\STATE $x(2i+1,2j+1)\gets w_b$
\ENDFOR
\ENDFOR
\end{algorithmic}
\end{algorithm}

\begin{figure*}[t]
\begin{minipage}{0.49\linewidth}
    \centering
    \centerline{\includegraphics[scale=0.55]{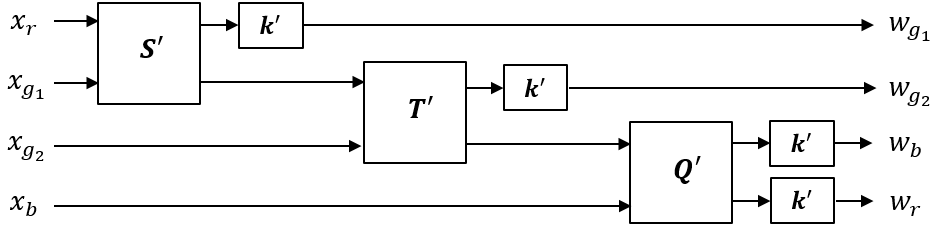}}
    \centerline{\footnotesize{(a)}}
\end{minipage}
\begin{minipage}{0.49\linewidth}
    \centering
    \centerline{\includegraphics[scale=0.55]{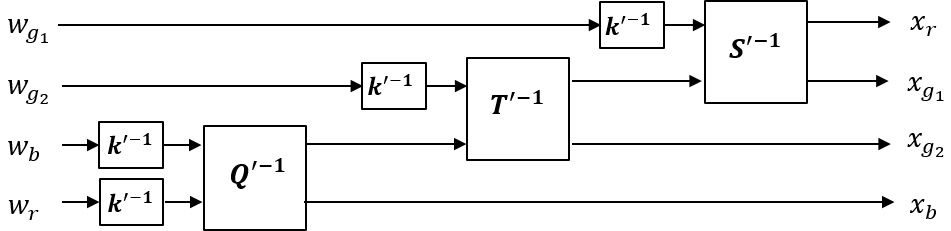}}
    \centerline{\footnotesize{(b)}}
\end{minipage}
\caption{``Sequential'' lossless white balance lifting scheme is an alternative to ``pyramid'' structure in \fref{fig:losslessWhiteBalance}. (a) Forward lifting steps. (b) Inverse lifting steps.}
\label{fig:losslessWhiteBalanceSequential}
\end{figure*}

There are several variations to the lossless white balance. For example, the scalar multiplication structure in \fref{fig:structureQ} is a pair-wise operation. The pairing and the ordering in \fref{fig:losslessWhiteBalance} can be changed. Alternatively, the scalar multiplication structure in \fref{fig:structureQ} can be applied in a sequential manner, as shown in \fref{fig:losslessWhiteBalanceSequential}. The weights need to be adjusted appropriately, obviously, but they can be solved by the system of log-weights similar to \eqref{eq:logwhitebalance}-\eqref{eq:equivWhiteBalance}. Experimentally, we found that differences in performance were negligible.

% Fig.4
%%%%%%%%%%%%%%%%%%%%%%%%%%%%%%%%%%%%%%%%

\section{Experimental Evaluation}
\label{sec:results}

We used raw sensor video sequences captured using FLIR Chameleon3\footnote{The captured raw video dataset can be downloaded from http://issl.udayton.edu/dataset} to evaluate the compression performance. The sensor resolution is $1024 \times 1280$, with 10 bit per pixel analog-to-digital converter. Of the eight sequences, one had 201 frames while the remainder had 299 frames. The frame rate of the test sequences was $75$fps. The sequences are diverse in contents, with indoor and outdoor scenarios---important for validating the merits of white balance in CFA compression under a diverse set of illumination colors.

\begin{table*}
\caption{Performance of lossless CFA compression with and without the proposed lossless white balance technique in terms of mega bits per second~(Mb/s). The coding gain of the white balance is computed as a percentage of the original CFA compression bit rate (i.e.~relative to the compression without the white balance). The bitrate of the uncompressed raw sensor data was 1465.852Mb/s.} \label{tab:bitrates}

\begin{center}
{\renewcommand{\arraystretch}{1.3}
\begin{tabular}{|c| c|c|c|| c|c|c|| c|c|c|}
\hline 
\multirow{3}{*}{Methods} & \multicolumn{3}{c||}{HEVC} & \multicolumn{3}{c||}{Motion JPEG2000} & \multicolumn{3}{c|}{Lossless JPEG} \\ \cline{2-10}
&  \multirow{2}{*}{Raw} & White Balanced & \multirow{2}{*}{Gain} &  \multirow{2}{*}{Raw} & White Balanced & \multirow{2}{*}{Gain} &  \multirow{2}{*}{Raw} & White Balanced & \multirow{2}{*}{Gain} \\ [-.05in]
&&(proposed)&&&(proposed)&&&(proposed)&\\\hline \hline
Color         & 2068.206& N/A & N/A  
             & 1762.173 & N/A &  N/A
             & 1879.035 & N/A & N/A  \\ \hline
Direct CFA    & 883.434 & 742.587 & {\bf 15.94\%}   
              & 698.541 & 680.174 & {\bf 2.63\%}  
              & 866.413 & 735.577 & {\bf 15.10\%}  \\ \hline
Demux         & 740.948 & 739.818 & 0.15\%   
              & 696.613 & 696.850 & -0.03\%  
              & 720.570 & 720.833 & -0.04\%      \\ \hline
MSST~\cite{Malvar12} & 739.528 & 723.082 & 2.22\%   
              & 696.743 & 683.335 & 1.92\%  
              & 722.979 & 711.339 & 1.61\%      \\ \hline
JPEG-XS~\cite{Richter2019Bayer}  & 732.055 & 722.801 & 1.26\%   
              & 690.925 & 683.959 & 1.01\%  
              & 717.970 & 711.861 & 0.85\%       \\ \hline
Mallat~\cite{Mallat98}  & 733.212 & 715.735 & 2.38\%   
              & 692.506 & 679.266 & 1.91\%  
              & 717.163 & 706.856 & 1.44\%       \\ \hline
CAMRA-A~\cite{lee2018camera} & 731.935 & 715.673 & 2.22\% 
              & 688.991 & 677.530 & 1.66\%
              & 715.872 & 706.413 & 1.32\%       \\ \hline
CAMRA-S~\cite{lee2020shift} & {\bf 727.297} & {\bf 714.144} & 1.81\% 
                            & {\bf 686.099} & {\bf 676.724} & 1.37\% 
                            & {\bf 713.740} & {\bf 706.084} & 1.07\% \\ \hline
\end{tabular}
}
\end{center}

\end{table*}

The lossless white balance technique proposed in this paper is agnostic to the choice of method used to estimate the illumination color $[\ell_r,\ell_g,\ell_b]$. In our implementation, the illumination colors $[\ell_r,\ell_g,\ell_b]$ is estimated from data using the gray world method (average of all red, green, and blue pixels in raw sensor data $x$, respectively). Although there are many white balance techniques available today~\cite{anderson1996proposal, ebner2007color, afifi2019color,chakrabarti2011color}, gray world method~\cite{land1983recent} has the advantage that the DC values of $z_{\gamma}$ and $z_{\beta}$ in \eqref{eq:lp} will be zero (because the average values of $z_r,z_g,z_b$ will equal). 

For evaluation, we follow the diagram in \fref{fig:system} to combine the proposed lossless white balance lifting scheme with existing CFA compression techniques. For the CFA encoder and decoder, we considered three ``baseline'' lossless compression schemes-----CFA sampled videos treated as gray images~(``Direct CFA''), and demultiplexed color videos (red, green, and blue sub-videos) of quarter resolution~(``Demux''). In addition, we implemented five state-of-the-art lossless CFA compression techniques---the macropixel spectral-spatial transformation~(``MSST'')~\cite{Malvar12}, the JPEG XS transformation~(``JPEG XS'')~\cite{Richter2019Bayer}, the Mallat wavelet packet transformation~(``Mallat'')~\cite{Mallat98}, the lossless camera-aware multi-resolution analysis~(``CAMRA-A'')~\cite{lee2018camera}, and shift-and-decorrelate CAMRA (``CAMRA-S'')~\cite{lee2020shift}. To yield an actual binary coding stream, we paired the baseline and state-of-the-art CFA compression techniques with three codecs---high efficiency video coding~(HEVC)~\cite{Sullivan12} (HM $16.20$ reference software in range extensions~(RExt) intra lossless mode~\cite{JCTVCHEVC}, %~\footnote{https://hevc.hhi.fraunhofer.de/trac/hevc/browser/tags}, 
which supports monochrome up to 16bits~\cite{Flynn16}), Motion JPEG2000~\cite{MPEG2000} ({\it OpenJPEG}~\cite{OpenJPEG}, 5 level LeGall $5/3$ wavelet transform~\cite{Adams00, MPEG2000} with $64\times64$ code blocks), and lossless JPEG~\cite{LSJPEG} (MATLAB ``imwrite''). For bench-marking purposes only, we also include another ``baseline''---full color video frames (``color''), compressed losslessly after demosaicking, color correction, white balance, and gamma correction applied. As is the standard practice for color images, it was compressed in YCbCr $4:4:4$ format.

We report the compression performance in Table \ref{tab:bitrates} in terms of megabits per second~(Mb/s), with and without the proposed white balance as well as the percentage bitrate reduction due to white balance. With the exception of Demux, all bitrates improved, though the rate of improvement depended on the CFA encoding method. For example, the best performing encoding method is ``CAMRA-S+Motion JPEG2000,'' which yielded 686.099 Mb/s (megabits per second) without white balance and 676.724 Mb/s with white balance---an improvement of 1.37\%. A similar performance gain due to lossless white balance is observed for CAMRA-A and MSST, which explicitly takes advantage of the luminance-chrominance decomposition, as well as Mallat, which performs spatial frequency decomposition~\cite{Lee17}. By comparison, improvement by lossless white balance was modest in JPEG-XS, which uses luminance-chrominance representation different from \eqref{eq:opponentcolor}. The post-demosaicking compression method (Color) proves to be an inefficient modality for lossless compression.

A large coding gain by white balance was also seen in the direct compression of CFA data. In particular, the performance of ``Direct CFA+Motion JPEG 2000'' with white balance was surprisingly competitive at 680.174 Mb/s, surpassing MSST and JPEG-XS. This is partly due to the fact that the neighboring red/green/blue pixels now have similar pixel values due to white balance. See \fref{fig:bayerPattern}(c). In addition, analysis in our previous work showed that the wavelet transform applied to CFA sampled data within Motion JPEG2000 implicitly performs the luminance-chrominance decomposition in \eqref{eq:opponentcolor}, similar to the ``Mallat'' CFA compression method~\cite{Lee17}. This is likely another contributing factor to the positive impact of lossless white balance on direct CFA compression.

The lossless white balance had negligible impact on the coding efficiency of Demux method, which neither takes advantage of the cross-color correlation nor performs CFA-level spatial decomposition. This is consistent with the analysis in Section \ref{sec:motivation} that although lossless white balance simply increases or decreases the relative dynamic range of the red/green/blue channels, the differential entropy of $x_r,x_{g_1},x_{g_2},x_b$ and $w_r,w_{g_1},w_{g_2},w_b$ remain the same. 

\section{Conclusion}
\label{sec:conclusion}

We proposed white balance using a scalar multiplication lifting scheme for lossless CFA image and video compression. Thanks to the integer operations in lifting the process are invertible, making it a perfectly lossless white balance operation. On its own, the changes in dynamic range stemming from white balance do not penalize nor benefit the coding efficiency. However, the white balance helps cancel out the highpass components in color channels, reducing the spatial bandwidth of the chrominance images. Hence a combination of lossless white balance and spatial decompositions (such as wavelet transform) improves the coding efficiency of state-of-the-art CFA compression methods. As future work, we plan to integrate the proposed lossless white balance directly into lossless CFA image and video compression (rather than treating white balance as a pre-processing step).

\bibliographystyle{IEEEtran}
\bibliography{refs}

\end{document}